\documentstyle[11pt,fleqn]{article}

\begin{document}
\title{ An Ising model in a magnetic field with a boundary}

\author{ Murray T. Batchelor, Vlad Fridkin and Yu-kui Zhou \\ \\
Department of Mathematics, School of Mathematical Sciences, \\
Australian National University, Canberra ACT 0200, AUSTRALIA}

\date{November 4, 1995}

\setlength{\mathindent}{1 mm}

\maketitle
\begin{abstract}
We obtain the diagonal reflection matrices for a recently introduced
family of dilute ${\rm A}_L$ lattice models in which the
${\rm A}_3$ model can be viewed as an Ising model in a magnetic field.
We calculate the surface free energy from the crossing-unitarity
relation and thus directly obtain
the critical magnetic surface exponent $\delta_s$ for $L$ odd and
surface specific heat exponent for $L$ even
in each of the various regimes.
For $L=3$ in the appropriate regime we obtain the Ising exponent
$\delta_s = -\frac{15}{7}$,
which is the first determination of this exponent
without the use of scaling relations.
\end{abstract}
\newcommand{\smat}[1]{\mbox{\small $\pmatrix{#1}$}}
\newcommand{\be}{\begin{eqnarray}}
\newcommand{\ee}{\end{eqnarray}}
\newcommand{\lam}{\lambda}
\newcommand{\hs}[1]{\hspace*{#1cm}}
\newcommand{\vs}[1]{\vspace*{#1cm}}
\newcommand{\no}{\nonumber}
\newcommand{\wt}[6]{#1\mbox{\small
 $\left(\matrix{#5&#4\cr#2&#3\cr}\biggm|\mbox{$#6$}\right)$}}
\newcommand{\Wf}[9]{W_{m\times n}\mbox{$\left(
   \matrix{#5&#8&#4\cr#9&&#7\cr #1&#6&#3\cr}\biggm|
       \mbox{$#2$}\right)$}}
\newcommand{\K}[5]{K_{#1}\biggl(\!\matrix{&#3\vs{-0.3}\cr\!\!
  #2\hs{-0.3}\vs{-0.3}&\cr&#4}\!\!\biggm|\!\mbox{$#5$}\biggr)}
\newcommand{\Kp}[4]{K_{+}\biggl(\matrix{#2\vs{-0.3}&\cr&
  \hs{-0.3}#1\vs{-0.3}\cr#3&}\!\!\biggm|\!\mbox{$#4$}\biggr)}
\newcommand{\T}{\mbox{\boldmath$T$}}

\def\cross#1#2#3#4#5{
\setlength{\unitlength}{0.01000in}%
\begin{picture}(30,33)(45,776)
\put( 45,795){\line( 1,-1){ 30}}
\put( 75,795){\line(-1,-1){ 30}}
\put( 43,777){\tiny #1}
\put( 56,763){\tiny #2}
\put( 69,777){\tiny #3}
\put( 56,789){\tiny #4}{#5}
\end{picture}}
\def\rightK#1#2#3#4{
\setlength{\unitlength}{0.0110in}%
\begin{picture}(15,30)(105,713)
\put(105,735){\line( 1,-1){ 15}}
\put(120,720){\line(-1,-1){ 15}}
\put(102,717.5){\tiny #1}
\put(117,707){\tiny #2}
\put(117,728){\tiny #3}{#4}
\end{picture}}
\vskip 1cm

There has been a recent growth of interest in the interaction-round-a-face
(IRF) formulation of lattice models in statistical mechanics in the
presence of a boundary \cite{BPO,Z,AK,BZ,ZBa,ZBb}. IRF models such as the
restricted solid-on-solid (RSOS) $A_L$ models \cite{ABF} and the
dilute RSOS $A_L$ models \cite{WNS,Roche} are particularly
attractive as their solution in terms of elliptic functions correspond to
off-critical extensions in which the elliptic nome $p$ measures the
deviation from the critical point $p=0$. We refer to these models here
as the $A_L$ models \cite{ABF} and the dilute $A_L$ models \cite{WNS}.
In the $A_L$ models
$p$ is temperature-like while for the dilute RSOS models $p$ is
temperature-like for $L$ even but is magnetic-like for $L$ odd.
In particular, the $A_3$ model, in the appropriate regime,
can be viewed as a critical Ising model in a thermal field.
On the other hand, the dilute $A_3$ model,
in the appropriate regime, can be viewed as a critical
Ising model in a magnetic field \cite{WNS}.
The singular part of the free energy of the $A_3$ model
yields the bulk Ising specific heat exponent $\alpha_b = 0$
\cite{ABF}. Consideration of the surface free
energy yields the known Ising surface specific heat exponent \cite{MW,bin}
$\alpha_s = 1$ \cite{ZBa}. On the other hand, the dilute $A_3$ model
provided a direct calculation of the bulk Ising magnetic exponent
$\delta_b=15$, first from the singular behaviour of the free
energy \cite{WNS} and later from a calculation of the
order parameters \cite{WPSN}.

In this article we consider the dilute $A_L$ models with open
boundaries and derive the surface free energy from which we obtain
the magnetic Ising surface exponent $\delta_s=-\frac{15}{7}$
in the appropriate regime of the $A_3$ model.
This is the first direct calculation
of this quantity without the use of scaling relations.

The dilute $A_L$ lattice models \cite{WNS,Roche}
are RSOS models with $L$ heights
built on the $A_L$ Dynkin diagram with a loop at each node.
The non-zero face weights of the off-critical dilute $A_L$ models
satisfy the star-triangle
relation \cite{Baxter} and are given by \cite{WNS}
\be
\wt Waaaau & = &
\frac{\vartheta_1({6\lambda-u})\vartheta_1({3\lambda+u})}{
    \vartheta_1({6\lambda})\vartheta_1({3\lambda})} \no \\
\hs{-1.2}& &\hs{-1.5}-\left(
 \frac{S(a+1)}{S(a)}\frac{\vartheta_4({2a\lambda-5\lambda})}{
                           \vartheta_4({2a\lambda+\lambda})}
 +\frac{S(a-1)}{S(a)}\frac{\vartheta_4({2a\lambda+5\lambda})}{
                            \vartheta_4({2a\lambda-\lambda})}\right)
\frac{\vartheta_1({u})\vartheta_1({3\lambda-u})}{
        \vartheta_1({6\lambda})\vartheta_1({3\lambda})} \no \\
\wt Waaa{a\pm 1}u & = & \wt Wa{a\pm 1}aau \;\; = \;\;
\frac{\vartheta_1({3\lambda-u})\vartheta_4({\pm 2a\lambda+\lambda-u})}{
      \vartheta_1({3\lambda})\vartheta_4({\pm 2a\lambda+\lambda})}\no \\
\wt W{a\pm 1}aaau & = & \wt Waa{a\pm 1}au \;\; =\;\;
\left(\frac{S(a\pm 1)}{S(a)}\right)^{1/2}
\frac{\vartheta_1({u})\vartheta_4({\pm 2a\lambda-2\lambda+u})}{
   \vartheta_1({3\lambda})\vartheta_4({\pm 2a\lambda+\lambda})} \no \\
\wt Wa{a\pm 1}{a\pm 1}au & = & \wt Waa{a\pm 1}{a\pm 1}u \no \\
\hs{-0.2}& = & \left(\frac{\vartheta_4({\pm 2a\lambda+3\lambda})
              \vartheta_4({\pm 2a\lambda-\lambda})}
           {\vartheta_4^2(\pm 2a\lambda+\lambda)}\right)^{1/2}
\frac{\vartheta_1({u})\vartheta_1({3\lambda-u})}{
     \vartheta_1({2\lambda})\vartheta_1({3\lambda})} \label{weights} \\
\wt Wa{a\mp 1}a{a\pm 1}u & = &
\frac{\vartheta_1({2\lambda-u})\vartheta_1({3\lambda-u})}{
    \vartheta_1({2\lambda})\vartheta_1({3\lambda})} \no \\
\wt W{a\pm 1}a{a\mp 1}au & = &
-\left(\frac{S(a-1)S(a+1)}{S^2(a)}\right)^{1/2}
\frac{\vartheta_1({u})\vartheta_1({\lambda-u})}{
      \vartheta_1({2\lambda})\vartheta_1({3\lambda})} \no \\
\wt W{a\pm 1}a{a\pm 1}au & = &
\frac{\vartheta_1({3\lambda-u})\vartheta_1({\pm 4a\lambda+2\lambda+u})}{
  \vartheta_1({3\lambda})\vartheta_1({\pm 4a\lambda+2\lambda})} \no\\
& & +\frac{S(a\pm 1)}{S(a)}
\frac{\vartheta_1({u})\vartheta_1({\pm 4a\lambda-\lambda+u})}{
  \vartheta_1({3\lambda})\vartheta_1({\pm 4a\lambda+2\lambda})} \no\\
\hs{-0.2}& = & \frac{\vartheta_1({3\lambda+u})
    \vartheta_1({\pm 4a\lambda-4\lambda+u})}
{\vartheta_1({3\lambda})\vartheta_1({\pm 4a\lambda-4\lambda})} \no \\
& & \hs{-1}+\left(\frac{S(a\mp 1)}{S(a)}
   \frac{\vartheta_1({4\lambda})}{\vartheta_1({2\lambda})}
-\frac{\vartheta_4({\pm 2a\lambda-5\lambda})}{
         \vartheta_4({\pm 2a\lambda+\lambda})} \right)
\frac{\vartheta_1({u})\vartheta_1({\pm 4a\lambda-\lambda+u})}{
   \vartheta_1({3\lambda})\vartheta_1({\pm 4a\lambda-4\lambda})}\no
\ee

\noindent
The crossing factors $S(a)$ are defined by
\be \hskip 10 mm
S(a) & = & (-1)^{\displaystyle a} \;\frac{\vartheta_1({4a\lambda})}{
           \vartheta_4({2a\lambda})}
\ee
and $\vartheta_1({u})$, $\vartheta_4({u})$ are standard elliptic
    theta functions of nome $p$
\be \hskip 10 mm
\vartheta_1(u)&=&\vartheta_1(u,p)=2p^{1/4}\sin u\:
  \prod_{n=1}^{\infty} \left(1-2p^{2n}\cos
   2u+p^{4n}\right)\left(1-p^{2n}\right)\label{theta1}  \\
\vartheta_4(u)&=&\vartheta_4(u,p)=\prod_{n=1}^{\infty}\left(
 1-2p^{2n-1}\cos2u+p^{4n-2}\right)\left(1-p^{2n}\right).
  \label{theta4}
\ee

Four different critical branches are defined by \cite{WNS}
\be \hskip 10 mm
&\mbox{branch {\it 1}}
\hs{1.8} 0<u< 3\lambda &\hs{0.8}\lambda ={\pi\over 4}{L\over L+1} \hs{0.7}
L = 2,3, \cdots \no\\
&\mbox{branch {\it 2}}
\hs{1.8} 0<u< 3\lambda  &\hs{0.8}\lambda ={\pi\over 4}{L+2\over L+1}\hs{0.7}
L = 3,4,\cdots    \no\\
&\mbox{branch {\it 3}}
\hs{0.5} -\pi+3\lambda <u<0 &\hs{0.8}\lambda ={\pi\over 4}{L+2\over
L+1}\hs{0.7} L = 3,4,\cdots \label{lam}\\
&\mbox{branch {\it 4}}
\hs{0.5} -\pi+3\lambda <u<0  &\hs{0.8}\lambda ={\pi\over 4}{L\over
L+1}\hs{0.7}L = 2,3, \cdots \no
\ee
This yields eight separate regimes, according to the sign of $p$.
The magnetic Ising point occurs in regime 2 with $\lambda=\frac{5\pi}{16}$.

The integrable boundary weights are represented by a triangular form
with three spins \cite{BPO,Z,AK}.
For the dilute models, in accordance with the adjacency condition of
the model, we can define
\be \hskip 30 mm
\K {}acbu=0 \hspace*{0.5cm}\mbox{unless $|a-b|=0,1$ and $|a-c|=0,1$}.
\ee
 These are to satisfy the boundary version of the star-triangle equation
(the reflection equation) \cite{Z,AK,Sklyanin}
\be \hskip 10 mm
&&\sum_{f,g}{\wt Wgcba{u-v}}{\K {}gcfu}{\wt Wdfga{u+v}}{\K {}dfev}
  \no\\
&&\hspace{0.5cm} =\sum_{f,g} {\K {}bcfv}{\wt Wgfba{u+v}}{\K {}gfeu
     }{\wt Wdega{u-v}} \label{BYBE}
\ee
ensuring the commutativity of the transfer matrix
$\mbox{\boldmath $T$}(u)$ defined by the elements
\be \hskip 10 mm
\langle\mbox{\boldmath $a$}|{\T}(u)|\mbox{\boldmath $b$}\rangle
&=&\sum_{\{c_0,\cdots,c_N\}} \Kp {c_0}{a_0}{b_0}u \left[
\prod_{k=0}^{N-1}
 \wt W{b_k}{b_{k+1}}{c_{k+1}}{c_k}{u} \right. \no \\
&& \left. \hspace{1cm}\times\wt W{c_{k+1}}{a_{k+1}}{a_k}{c_k}{u} \right]
 \K -{\;c_N}{a_N}{b_N}u    \label{openT}
\ee
where $\mbox{\boldmath $a$}=\{a_0,a_1,\cdots,a_N\}$ and
$\mbox{\boldmath $b$}=\{b_0,b_1,\cdots,b_N\}$ and
\be \hskip 30 mm
\K -acbu &=& \K {}acbu \\
\Kp acbu &=& \sqrt{\frac{S(a)^2}{S(b) S(c)}} \K {}acb{3\lambda-u}.
\ee
Note that the formulation of the reflection equation and transfer matrix
given in \cite{BPO} is a special case of the general one given here.
The above formulation does not incorporate crossing symmetry of the
bulk weights and is applicable to $A_n^{(1)}$, $n\ge 2$ \cite{BFKZ} for which
a crossing symmetry does not exist.

Another important relation is the boundary
crossing relation, which here reads
\be \hskip 30 mm
\sum_{c}\sqrt{S(c) \over S(a)}
\wt Wabcd{2u+3\lambda} \K {}cdb{u+3\lambda}\nonumber \\\qquad\qquad=
{\vartheta_1(
 2u+5\lambda) \vartheta_1(2u+6\lambda)
 \over \vartheta_1(2\lambda)\vartheta_1(3\lambda)} \K {}adb{-u}\;.
\label{Kcrossing}
\ee

It can be easily seen that the reflection equation reduces to five
nontrivial equations with three distinct forms for diagonal boundary
weights.  These three distinct
forms correspond to the generic equations in the higher rank
$B^{(1)}_n, A^{(2)}_n$ cases, which have been solved in \cite{BFKZ}.  The
diagonal solution we find is
\be \hskip 30 mm
\K {}{\;a\pm 1}aau &=&\frac{\vartheta_4(\pm 2a\lambda+\xi-u)}
      {\vartheta_4(\pm 2a\lambda+\xi+u)}\; g_a(u) \label{K1}\\
\K {}aaau &=&
 \frac{\vartheta_1(-\lambda+\xi-u)}{\vartheta_1(-\lambda+\xi +u)} \; g_a(u)
\label{K2}\ee
where
\[ \hskip 30 mm
g_a(u) =
\vartheta_1(\xi+u)\vartheta_1(-\lambda+\xi+u)\vartheta_4(2a\lambda+\xi+u)
\vartheta_4(-2a\lambda+\xi+u).
\]
Here the parameter $\xi$ takes the values
$\xi = -\lambda /2$ mod ($\ell\pi/2 + m\pi\tau/2$)
where $\ell$ and $m$ are integers and $g_a(u)$ has
been fixed by crossing symmetry, otherwise $g_a(u)$ may
be taken arbitrarily.  The freedom in $\xi$ due to the quasi-periodicity
of the elliptic functions gives four distinct solutions.
All of these satisfy crossing symmetry
up to a gauge factor.  As for the bulk weights, height reversal symmetry
is broken by the boundary weights for $L$ odd, ensuring that the
nome $p$ can again be regarded as a magnetic field.

At criticality the known $K$-matrix solutions for the $A_2^{(2)}$
vertex model \cite{MN} are recovered via the usual passage from
face weights to vertex weights. The integrable boundary
weights of the dilute O($n$) loop model \cite{YB} are also recovered in
this limit after transforming to the diagonal orientation \cite{ZBa}.

The fusion procedure has been applied to the dilute $A_L$ models,
resulting in the construction of both $su(2)$ \cite{Zhou}
and $su(3)$ \cite{ZPG} fused face weights from the face weights
given in (\ref{weights}). The $su(2)$ fusion rule provides the functional
relation \cite{Zhou}
\be \hskip 30 mm
\T(u)\T(u+3\lambda)=\mbox{\boldmath $f$}(u)
   +\T^{(2)}(u)
\ee
where $\T^{(2)}(u)$ is the transfer matrix of the fused
model with fusion level 2.
For periodic boundary conditions the matrix function
$\mbox{\boldmath $f$}(u)$ is given by
\be \hskip 30 mm
\mbox{\boldmath $f$}(u)=[\rho(u)]^N  \mbox{\boldmath $I$}
\ee
where $\mbox{\boldmath $I$}$ is the identity and
\be \hskip 30 mm
\rho(u)={\vartheta_1(2\lambda-u) \vartheta_1(3\lambda-u)
   \vartheta_1(2\lambda+u) \vartheta_1(3\lambda+u)
 \over \vartheta_1^2(2\lambda)\vartheta_1^2(3\lambda)}.
\ee

The fusion procedure can be carried out in a similar manner for the
open boundary system, as has been done already for a number of IRF models
\cite{BPO,Z,ZBa,ZBb}. It is necessary to fuse the
boundary face weights (\ref{K1})-(\ref{K2}). We find the matrix
function $\mbox{\boldmath $f$}(u)$ to be diagonal with
element $(c,d)$ given by
\be \hskip 30 mm
f(u)_{c,d}
=\omega^-_c(u)\omega^+_d(u)\rho^{2N}(u)/\rho(2u) \label{f}
\ee
where the boundaries contribute the factors $\omega^-_c(u)$
and $\omega^+_d(u)$, with
\be \hskip 10 mm
\omega_c^-(u)&=&\sum_{b} \sqrt{S(b)\over S(c)} \;
  \wt Wccbc{2u+3\lambda}\K -bcc{u+3\lambda}\K -cccu \\
\omega^+_d(u)&=& \sum_{b} \sqrt{S(d)\over S(b)} \;
  \wt Wbddd{3\lambda-2u}\Kp ddd{u+3\lambda}\Kp bddu.
\ee
Height $c$ ($d$) is located on the right (left) boundary.

We now turn to the calculation of the magnetic surface exponents
$\delta_s$ from the surface free energy. The finite-size corrections
to the transfer matrix $\T(u)$ are contained in $\T^{(2)}(u)$.
According to the spirit of \cite{Z,BZ,ZBa,ZBb}
the boundary crossing unitarity relation
\be \hskip 30 mm
T(u)T(u+3\lambda)=f(u)_{c,d}
\ee
for the eigenvalues is sufficient to determine both the bulk and
surface free energies.
The two contributions can be separated out by setting
$T(u) = T_b(u) T_s(u)$ and defining
$T_b = \kappa_b^{2N}$ and $T_s = \kappa_s$. The free energies
per site then follow as $f_b(u) = -\log \kappa_b(u)$ and
$f_s(u) = -\log \kappa_s(u)$. Thus for the bulk contribution we have
\be \hskip 30 mm
\kappa_b(u) \kappa_b(u+3\lambda)= \rho(u).
\ee
This relation has already been used to determine the bulk free energy
\cite{WNS,WPSN} via the inversion relation method \cite{Baxter,Baxter:82}.
The critical behaviour as $p\rightarrow 0$, obtained by use of the
Poisson summation formula \cite{Baxter}, is \cite{WNS,WPSN}
\be \hskip 30 mm
f_b \sim \left\{ \begin{array}{ll}
                 p^{1 + 1/\delta_b} & \mbox{$L$ odd} \\
                 p^{2 - \alpha_b}   & \mbox{$L$ even}
             \end{array} \right.
\ee
where the values of $\delta_b$ and $\alpha_b$ are listed for the
different regimes in Table 1.

To determine the surface exponents $\delta_s$ and $\alpha_s$ it is
sufficient to
consider only the relevant contribution to the surface free energy,
which does not involve the explicit form of the boundary weights.
By making use of the boundary crossing relation and after
taking a convenient normalisation, the appropriate contribution
is obtained by solving
\be \hskip 30 mm
\kappa_s(u) \kappa_s(u+3\lambda)=
{\vartheta_1(5\lambda-2u) \vartheta_1(6\lambda-2u)
   \vartheta_1(5\lambda+2u) \vartheta_1(6\lambda+2u)
 \over \vartheta_1^2(5\lambda)\vartheta_1^2(6\lambda)}.
\label{ir}
\ee
We solve this relation under the same analyticity assumptions
as for the bulk calculation \cite{WNS,WPSN}. Specifically, we set
$p=e^{-\epsilon}$, make the relevant conjugate modulus transformations,
Laurent expand $\log \kappa_s(u)$ in powers of $\exp(-2\pi u/\epsilon)$
and match coefficients in (\ref{ir}). In regimes 1 and 2 we obtain
\setlength{\mathindent}{2mm}
\be
f_s = - 4 \sum_{k=1}^{\infty}
\frac{
\cosh (\pi \lambda k/\epsilon) \,
\cosh [(11 \lambda-\pi) \pi k/\epsilon] \,
\sinh (2 \pi u k/\epsilon) \,
\sinh [2(3\lambda-u)\pi k/\epsilon]}
{k\, \sinh (\pi^2 k/\epsilon) \,
\cosh (6 \pi \lambda k/\epsilon)}.
\ee
For regimes 3 and 4 the lhs of relation (\ref{ir}) needs
to be modified to $\kappa_s(u) \kappa_s(u+3\lambda-\pi)$ for the
appropriate analyticity strip, with end result
\be
f_s = 4 \sum_{k=1}^{\infty}
\frac{
\cosh (\pi \lambda k/\epsilon) \,
\cosh [(11 \lambda-\pi) \pi k/\epsilon] \,
\sinh (2 \pi u k/\epsilon) \,
\sinh [2(\pi\!-\!3\lambda+u)\pi k/\epsilon]}
{k\, \sinh (\pi^2 k/\epsilon) \,
\cosh [2(\pi\! -\! 3 \lambda) k/\epsilon]}.
\ee
Application of the Poisson summation formula yields
\be \hskip 30 mm
f_s \sim \left\{ \begin{array}{ll}
                 p^{1 + 1/\delta_s} & \mbox{$L$ odd} \\
                 p^{2 - \alpha_s}   & \mbox{$L$ even}
             \end{array} \right.
\ee
as $p\rightarrow 0$,
where the values of $\delta_s$ and $\alpha_s$ are listed for the
various regimes alongside the bulk values in Table 1.
As in the bulk case, these
exponents are magnetic for $L$ odd and thermal for $L$ even.

For $L=3$ in regime 2 we obtain the value $\delta_s=-\frac{15}{7}$.
This result is in agreement with the prediction for the
two-dimensional Ising model in a magnetic field using the
scaling relations between bulk and surface
exponents \cite{BH,B,F,bin}\footnote{Actually there is a misprint
in the tabulated result
for $\delta_s$ in Binder's review \cite{bin}.} The negative value
indicates that the surface magnetisation diverges towards infinity as
the magnetic field goes to zero. Thus a surface magnetisation in
zero field does not exist. Similar behaviour is known to occur in
the spherical model \cite{bin}. As expected, the Ising specific heat
exponents $\alpha_b = 0$ and $\alpha_s=1$ are recovered
in regimes 1 and 4 for $L=2$.

\hskip -10 mm \begin{table}
\caption{Magnetic and thermal critical bulk and surface exponents of the
dilute $A_L$ models.}
\begin{itemize}
\item[]\begin{tabular}{@{}*{7}{c}}
regime&1&2&3&4&&\cr
\cr
$\delta_b$&$\frac{3L}{L+4}$&$\frac{3(L+2)}{L-2}$&$\frac{L-2}{3(L+2)}$
&$\frac{L+4}{3L}$&$L$ odd\cr
\cr
$\delta_s$&$\frac{-3L}{L-2}$&$\frac{-3(L+2)}{L+4}$&$\frac{L+4}{L-2}$
&$\frac{L+4}{L-2}$&$L$ odd\cr
\cr
$\alpha_b$&$\frac{2(L-2)}{3L}$&$\frac{2(L+4)}{3(L+2)}$&$\frac{-2(L+4)}
{L-2}$&$\frac{-2(L-2)}{L+4}$&$L$ even\cr
\cr
$\alpha_s$&$\frac{2(2L-1)}{3L}$&$\frac{2(2L+5)}{3(L+2)}$&$\frac{-6}
{L-2}$&$\frac{6}{L+4}$&$L$ even\cr
\end{tabular}
\end{itemize}
\end{table}

\section*{Acknowledgements}
It is a pleasure to thank Atsuo Kuniba for helpful
discussions. MTB and YKZ are supported by the Australian Research Council
and VF is supported by an Australian Postgraduate Research Award.

\end{document}